\documentclass[twocolumn,aps,pre,showpacs,amsmath,amssymb,showpacs]{revtex4}

\usepackage{graphicx}
\usepackage{dcolumn}
\usepackage{bm}

\begin{document}

\title{Splashing of liquids: interplay of surface roughness with surrounding gas}
\author{Lei Xu}
\author{Loreto Barcos}
\author{Sidney R. Nagel}
\affiliation{The James Franck Institute and Department of Physics,
The University of Chicago, 929 East 57th St., Chicago, Illinois
60637}
\date{\today}

\begin{abstract}
We investigate the interplay between substrate roughness and
surrounding gas pressure in controlling the dynamics of splashing
when a liquid drop hits a dry solid surface.  We associate two
distinct forms of splashing with each of these control parameters:
prompt splashing is due to surface roughness and corona splashing is
due to instabilities produced by the surrounding gas.  The size
distribution of ejected droplets reveals the length scales of the
underlying droplet-creation process in both cases.
\end{abstract}

\maketitle

\noindent\textbf{I. Two contributions to splashing}
\\
\\
A liquid drop hitting a surface will often splash, rupturing into
many tiny droplets. Although splashing is ubiquitous and important
in many applications~\cite{1,2,3}, the control parameters governing
its occurrence have not yet been fully explored.  Impact velocity,
surface tension, viscosity and substrate roughness have long been
known to be important~\cite{4,5,6}.  However only recently was it
discovered that splashing can be completely eliminated on smooth
surfaces simply by lowering the surrounding gas pressure~\cite{7}.
In this paper we focus on the role that surface roughness plays in
producing a splash. Given the discovery that the gas pressure can,
by itself, cause splashing, it is apparent that only by removing the
surrounding gas can the other control parameters, such as surface
roughness, be investigated in isolation.  In order to make progress,
we therefore eliminate all vestiges of the corona splash by working
at low gas pressures.  This insures that the effects of surface
roughness are unperturbed by the effects of the surrounding gas.

Two types of splashing are well-known in the literature~\cite{8}:
``corona'' splashing, where a corona forms and subsequently ruptures
and ``prompt'' splashing, where droplets emerge at the advancing
liquid-substrate contact line.  Although the two phenomena are
distinct there has been no fundamental understanding of their
separate causes.  The results reported here suggest a simple
explanation: corona splashing is due to the presence of surrounding
gas and occurs above a critical gas pressure and prompt splashing
occurs on rough surfaces even in the absence of air.  By studying
the ejected-droplet size distribution of these two forms of
splashing, we can determine the characteristic length scale of the
droplet-creation process in each case.  This corroborates our
interpretation of how the splash is formed.

Our previous experiment~\cite{7} clearly showed that surrounding gas
causes the corona splash on smooth surfaces.  Here we study the
effect of another control parameter: surface roughness.  In order to
do this cleanly, we must lower the gas pressure until the effects of
air are negligible.  We achieved this by doing an experiment in a
helium atmosphere at 13kPa pressure, much lower than the threshold
pressure ~\cite{7} for the surrounding gas to produce a corona
splash. In this situation, therefore, splashing is caused entirely
by surface roughness.

In our experiments, we use ethanol which has density $\rho = 0.789g/cm^3$, viscosity $\nu = 1.36cSt$ and surface tension $\sigma = 22mN/m$.  We filmed drops, of diameter $D = 3.4 \pm 0.1$ mm, released inside a transparent vacuum chamber from a
nozzle at a height $95$ cm above a rough substrate. The impact speed
of the drop, $V_0 = 4.3 \pm 0.1m/s$, was determined by analyzing the
drop position in subsequent frames of each movie. Rough substrates
were obtained by using high-quality sandpapers uniformly coated with closely
packed particles (\emph{microcut-paper-discs}$^{\scriptsize\textregistered}$, Buehler
Ltd.).  The roughness, $R_a$, defined as the average
diameter of the particles, was varied between  $3\mu m$ and $78\mu
m$.  Clean glass microscope slides were used as the smooth
substrates.  The gas pressure, $P$, was varied. Ethanol wets our
substrates and does not rebound after hitting the surface \cite{9}.
We also note that although ethanol wets the substrate (zero degree
static contact angle), the profile of the liquid film during
expansion has a finite thickness at the edge, instead of decreasing
smoothly to zero. All experiments were done at constant temperature,
$23.6\,^{\circ}\mathrm{C}$

As we increase the roughness, $R_a$, we see an evolution in behavior
as shown in Fig. 1.  We compare the splash created solely by surface
roughness at low pressure with the splash created when air is also
present.  In each panel the top row is at low pressure, $P = 13kPa$,
and the bottom row is at atmospheric pressure, $P = 100kPa$.  Fig.1a
shows the result for a smooth surface: at low pressure there is only
a liquid film expanding smoothly on the substrate with no splashing
whereas at atmospheric pressure there is a corona that breaks up
into many small droplets.  This shows unambiguously that air causes
corona splashing.  Fig. 1b shows the result for a small amount of
roughness, $R_a = 5\mu m$.  In this case, at low $P$ there are two
regimes: an early stage with prompt splashing which is followed by a
peaceful regime where no splashing occurs.  At atmospheric pressure,
there is a single regime throughout the expansion that resembles
corona splashing except that it is not as symmetric as when the
surface is completely smooth.  Thus, for roughness $R_a = 5\mu m$ ,
there is a clear difference as $P$ is increased, showing that
surrounding gas can still be important at atmospheric pressure.  For
large roughness, $R_a = 78\mu m$, we see typical prompt splashing,
where droplets are ejected from the expanding contact line during
the entire expanding process for both low and atmospheric pressures.

\begin{figure}
\begin{center}
\includegraphics[width=3.1in]{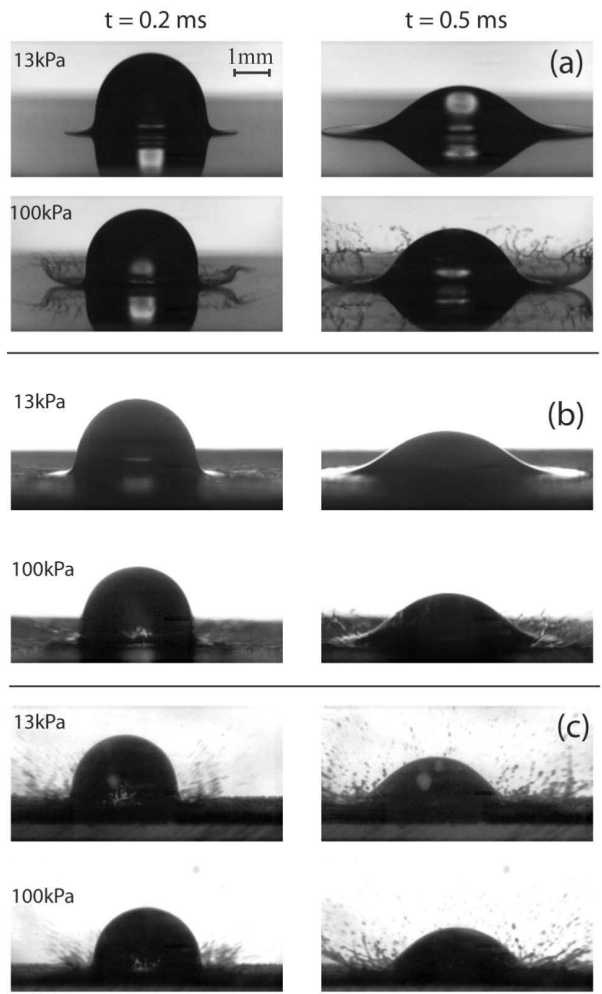}%
\caption{ Photographs of splashing as a function of gas pressure and
surface roughness.  The left and right columns are $0.2ms$ and
$0.5ms$ from the time of impact. For each value of surface
roughness, the top panel is at a low pressure, $P= 13kPa$ and the
bottom panel is at atmospheric pressure, $P= 100kPa$. (a) Splash on
a smooth substrate which is a clean microscope slide. (b) Splash on
a substrate with roughness $R_a = 5\mu m$. (c) Splash on a substrate
with roughness $R_a = 78\mu m$.}
\end{center}
\end{figure}

In summary, there are two contributions to splashing: a corona part
caused by gas and a prompt part caused by surface roughness.  At low
pressures, for small roughness ($R_a = 5\mu m$), there is no corona
and there is only a small amount of prompt splashing produced at the
beginning.  This prompt splashing disappears at later times.   For
large roughness ($R_a = 78\mu m$), splashing at the advancing
contact line is produced throughout the entire duration of the
expansion.  At atmospheric pressure there is a transition as the
roughness is increased: the corona splash dominates at small surface
roughness and the prompt splash dominates at large $R_a$.

These results indicate that in the absence of air, splashing is
caused when the expanding liquid film, of thickness $d$, becomes
destabilized by surface roughness; but if the roughness is too small
or the film is too thick then no splashing will occur.  Initially the expanding film thickness $d$ is of molecular size and
increases in thickness during expansion as liquid is added to the film.  Thus for
small roughness ($R_a =5\mu m$), splashing occurs immediately after
impact and is followed by a quiescent stage as the film becomes much
thicker than the roughness. For large roughness ($78\mu m$),
splashing continues throughout the film expansion since $d$ never
grows large enough to be unperturbed by the roughness.  From the
photographs, we estimated the liquid film thickness, $d$, at the
point where splashing stops for $R_a =5\mu m$ and found $d\sim 50\mu
m$.  This suggests the following criterion for prompt splashing:
\begin{equation} \label{eq1}
\frac{R_a}{d}=C(We,Re)
\end{equation}
\noindent where $C(We,Re)$ is a dimensionless number depending on
Weber number $We=\rho V_0^2 D/\sigma$ and Reynolds number
$Re=\rho V_0 D/\mu$.  For the impact conditions in Fig.1, $We\approx2400$ and
$Re\approx11500$, we conclude $C\sim0.1$.  Further studies are necessary to establish the dependence of  $C$ on $We$ and $Re$.  For example, at sufficiently low velocity we might not
expect to see any splash at all.  We emphasize that $C(We,Re)$ can only
be measured accurately when the effect of gas pressure is negligible
since otherwise the corona component will contaminate the results.
This shows the importance of separating the two types of splashing
by working at small $P$.
\\
\\
\noindent\textbf{II. Size distribution of prompt splashing}
\\
\\
After impact, the liquid breaks up into many tiny droplets.  The
size distribution of these emitted droplets, $N(r)$, may retain an
imprint of the droplet creation process and thus provide a clue to
the mechanism initiating the interfacial instability.  Here we
report $N(r)$ for prompt splashing on rough substrates.

\begin{figure}[!t]
\begin{center}
\includegraphics[width=3in]{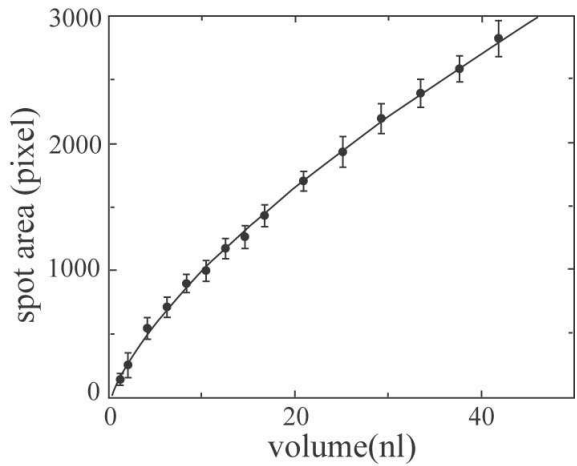}%
\caption{Calibration curve of stain area as function of droplet
volume. Error bars come from the fluctuation of stain areas for the
same size droplet. The fitting curve is: $y=233*x^{0.67}-119$. y is
the stain area in units of pixels, and x is the volume of the drop
in units of nanoliters.}
\end{center}
\end{figure}

\begin{figure}[!t]
\begin{center}
\includegraphics[width=3.1in]{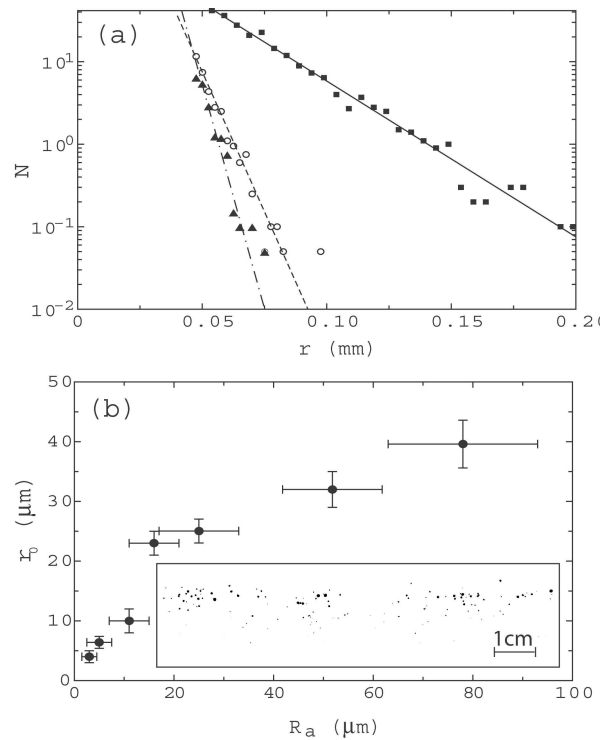}%
\caption{The distribution of ejected drops in a prompt splash on
rough substrates.  (a) $N(r)$ versus $r$ for splashes on substrates
with three values of roughness:  $R_a=16\mu
m$({\tiny$\blacksquare$}), $R_a=5\mu m(\circ)$, and $R_a=3 \mu
m(\blacktriangle)$. The exponential fitting functions are:
{\large---} , exp(-r/0.023); {\large{-}{-}{-}} , exp(-r/0.006); and
{--}$\cdot${--} , exp(-r/0.004). (b) The decay constant, $r_0$, of
the exponential decay in $N(r)$, as a function of substrate
roughness $R_a$.  For small values of roughness, the decay constant
is approximately linear in the roughness.  At large roughness, the
decay constant saturates. The sizes of particles on sandpaper are
randomly distributed around an average value, $R_a$. The fluctuation
of particle sizes gives us $R_a$ error bars; and the standard
deviation from exponential fit gives the $r_0$ error bars. Inset of
(b) shows one sheet of paper with ink spots produced by $R_a = 25
\mu m$. Most spots are randomly distributed within a narrow band.}
\end{center}
\end{figure}

We measure the size of emitted droplets by adding a small amount of
ink (\emph{Sanford}$^{\scriptsize\textregistered}$ black stamp pad
inker) to the ethanol, at the volume ratio 1:6 (ink:ethanol).  The
mixed liquid has the following material properties: density: $ \rho
= 0.833 \pm 0.002g/cm^3$, viscosity: $\nu =3.4 \pm 0.2 cSt$, surface
tension: $\sigma = 23 \pm 3 mN/m$. ($\sigma$ is measured both in air
and helium atmosphere under low pressure.  Both cases give the same
result, $23 \pm 3 mN/m$.)  Except for the viscosity, which increases
by a factor of 2.5, these values are close to those for pure
ethanol.  We have also checked repeatedly with high-speed video and
found that the splashing pictures look very similar with and without
ink, therefore we made sure that the addition of ink does not change
splashing qualitatively. We surround the impact point with a
cylinder of diameter $8.89 \pm 0.02 cm$ rolled from a sheet of white
paper. After splashing, the droplets hit the cylinder, leaving
stains.

With careful calibration, we convert the sizes of the ink
spots on the paper to the sizes of the ejected droplets.  Our calibrations work
well for drops with radius larger than $r=50\mu m$.  Below that
radius, our resolution is inadequate to obtain a reliable calibration.
We used a $0.5\mu l$ syringe to deliver tiny droplets of the ethanol/ink fluid with
known volume to white paper.  By measuring the area of the stains created by these
droplets we obtained the calibration curve in Fig.2.  We checked the
effect of droplet velocity on the stain area and found no significant
effect within our experimental accuracy.   By checking the shape of
a stain, we can ascertain whether that spot was caused by a single
ejectile or by two separate drops that landed in overlapping
locations.  We found less than $4\%$ overlapping stains.  We
excluded them from our distribution curves.

We first determine $N(r)$ for prompt splashing on rough substrates.
To ensure that these distributions are due solely to prompt
splashing, we performed the measurements at low pressure.  Fig. 3a
shows $N(r)$ for several values of roughness, $R_a$.  To obtain good
statistics, each distribution is an average over ten to twenty
experiments taken under the same conditions.  We added the number of
spots found in all the experiments and then divided the total counts
by the number of experiments. The straight lines in the figure
indicate that $N(r)$ decays exponentially with a characteristic
decay length, $r_0$:
\begin{equation} \label{eq3}
N \sim exp (-r / r_0)
\end{equation}
\noindent  Fig. 3b shows that the decay length of these lines,
$r_0$, increases with increasing $R_a$.  At small $R_a$, the decay
length is close to the value of the roughness: $r_0 \approx R_a$.
At large roughness, this relationship breaks down as  $r_0$ appears
to saturate at a constant value.

This behavior is consistent with the prompt splashing criteria, Eq. 1.  When the roughness is small, the drop stops splashing at a film thickness determined by $R_a$.  This sets the correlation
between the droplet decay length and roughness: $r_0 \approx R_a$.
However, when the roughness is too large, the drop never stops
splashing, and $r_0$ can only increase up to the maximum thickness
of the expanding liquid film.   From the
photographs we estimate thickness $d\sim 100\mu m$.  Fig. 3b shows that the saturation occurs at
approximately $40\mu m$, which is about the same order of magnitude as the
thickness of the liquid film at its terminal position.

\begin{figure}
\begin{center}
\includegraphics[width=3.1in]{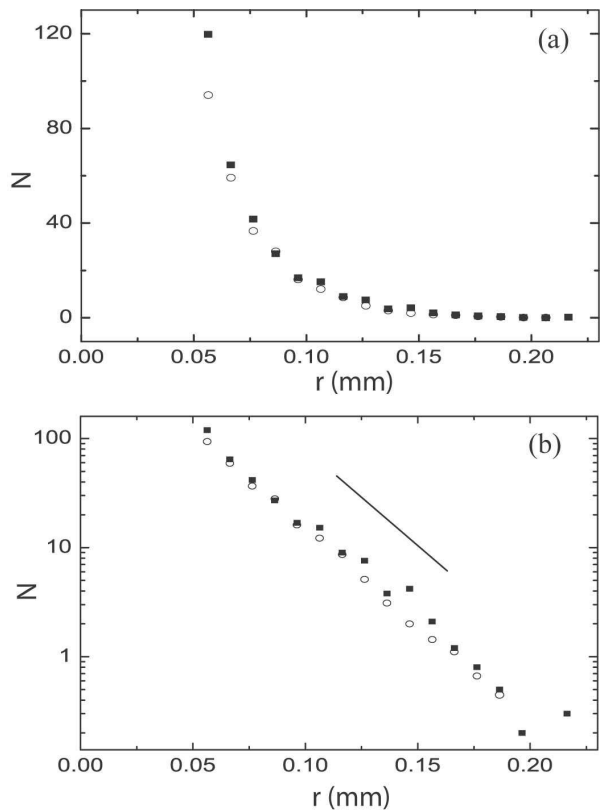}%
\caption{Size distribution of different impact velocities. For a
fixed roughness $R_a = 25 \mu m$, two impact velocities were tested:
$V_0 = 4.3 \pm 0.1 m/s(\circ)$ and $V_0 = 5.2 \pm 0.1 m/s$({\tiny
$\blacksquare$}). (a) shows N(r) in linear-linear scale. It's clear
that the higher velocity case produces about 20\% more splash. (b)
plots the same data in log-linear format. The two data sets now seem
much closer only because of the log-linear way we plot them. We can
fit both curves with the same functional form: $A \cdot
exp(-r/0.025)$(the solid line), by only varying the amplitude, A.
This implies that $r_0=0.025mm$ is independent of impact velocity.}
\end{center}
\end{figure}

How robust is the decay length, $r_0$, with respect to variations of
the impact velocity $V_0$, drop size $R$, or surface tension
$\sigma$?  In Fig. 4 we show the size distribution for two different
impact velocities: $V_0 = 4.3 \pm 0.1 m/s$($We=2155$, $Re=3648$) and
$V_0 = 5.2 \pm 0.1 m/s$($We=3151$, $Re=4412$). Fig. 4(a) shows that
there is significantly more splashing at higher impact velocity.
Fig. 4(b) plots the same data in log-linear format and shows that
both curves can be fitted by straight lines with the same decay
length, $r_0$. This implies that $r_0$ is independent of $V_0$.
Further experiments addressing other parameters should be done in
the future.  Although the absolute value of $r_0$ might change with
those parameters, we expect that the two general features: (1) the
distribution decays exponentially and (2) $r_0$ increases with
surface roughness, would still be valid.
\\

\noindent\textbf{III. Size distribution of corona splashing}
\\

\begin{figure}
\begin{center}
\includegraphics[width=3.1in]{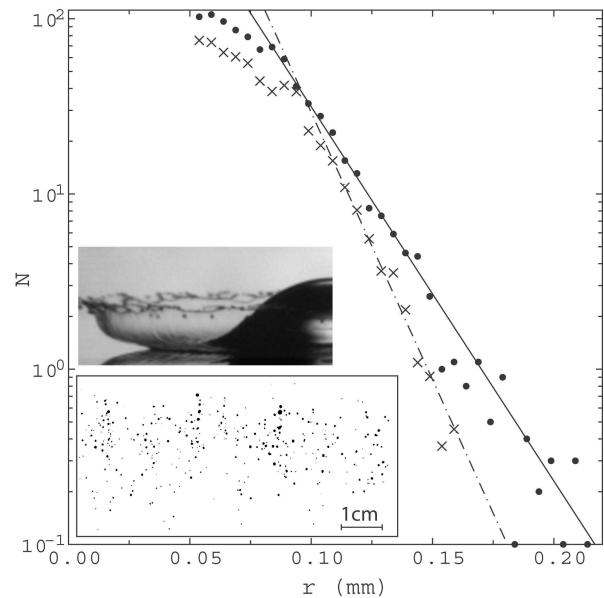}%
\caption{Size distribution of ejected droplets in a corona splash at
high pressure. Upper inset shows the corona film before it breaks
up.  From this picture we can estimate the film thickness to be $20
\sim 40 \mu m$ by measuring the thickness of the corona rim.  (Although rim should be
thicker than the film, we assume that the
film and the rim are approximately the same size.)  Lower inset is a reproduction of the
sheet of paper with the ink spots showing that the ejected droplets
hit the paper at random locations over a large area.  Main panel shows
the number of droplets of a given size per impact, $N(r)$, versus droplet
radius, $r$, for a corona splash at two pressures: $P=100$kPa
($\bullet$) and $P=80$kPa ($\times$). The exponential fitting
functions are respectively: {\large---} , exp(-r/0.020); and
{--}$\cdot${--} , exp(-r/0.014). }
\end{center}
\end{figure}

\begin{figure}
\begin{center}
\includegraphics[width=3.1in]{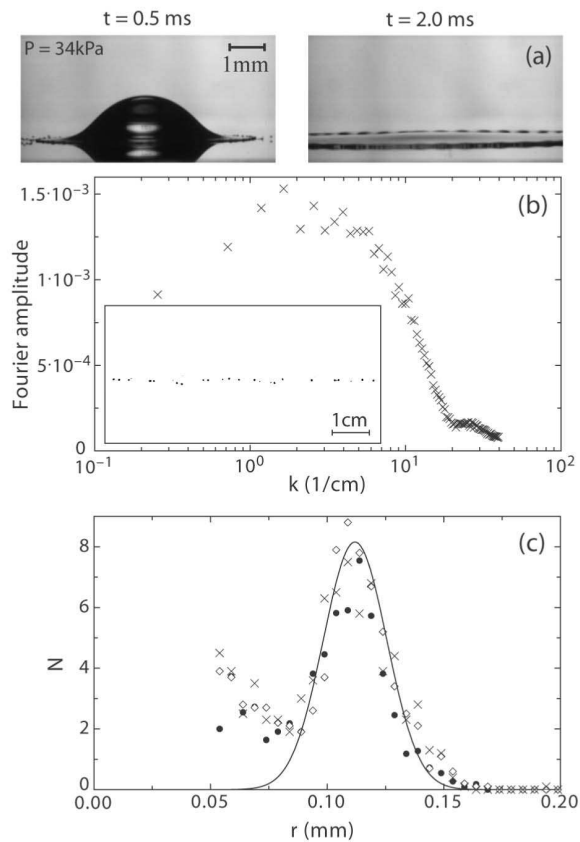}%
\caption{A splash just above the threshold pressure, $P_T$. (a)
Images of a drop splashing show the ejection of droplets and the
undulations in the expanding rim. The times shown are measured with
respect to the time of impact. (b) Inset is a reproduction of the
sheet of paper showing that the droplets hit the paper in a
well-defined horizontal line.  Main panel of (b) shows the Fourier transform
of the lateral positions of the spots in the inset.  The peak indicates a
well-defined spacing between the ejected droplets at $P_T$. (c) The
number of droplets at fixed sizes, $N(r)$, is plotted versus droplet
radius, $r$, at pressures slightly higher than $P_T$: P
34kPa($\bullet$), P= 37 kPa($\circ$) and P = 39kPa($\times$). The
solid line is a Gaussian fit centered at $r_0 = 0.11$mm. The peak in
$N(r)$ shows that the average size of the droplets is approximately the
size of the undulations at the rim of the expanding film.}
\end{center}
\end{figure}

We have also performed measurements of the size distributions
produced in a corona splash. This allows us to compare prompt and
corona splashing. Fig. 5 shows the results of splashing on a smooth
surface at two pressures, $100$kPa and $80$kPa, both in the high
pressure regime. The inset shows the spots created by the droplets
ejected from a single splash.  The  spots are randomly distributed.
The main panel shows the distribution of droplet sizes, $N(r)$, for
the two pressures. We again see an exponential distribution at large
radius $r$, $N \sim exp (-r / r_0)$, indicating the existence of a
characteristic length scale. The values we find for $r_0$ (given in
the caption of Fig. 5) are comparable to the corona thickness which we
estimated from the movies to be between $20$ and $40 \mu m$.  (The
corona thickness was estimated by measuring the edge of the corona
in pictures similar to the one shown in the inset to Fig. 5.)  This
suggests that the corona thickness determines $r_0$.  That $r_0$
decreases as the pressure decreases suggests that lower pressure
leads to less splashing and a thinner corona. The data in Fig. 5
shows that the size-distribution data can probe the slight
differences in the corona formed at the two pressures.

Both prompt and corona splashing have exponential distributions for
$N(r)$.  Although there is a characteristic length scale in both
cases, the control parameters governing the length scales are
different: roughness in Fig. 3 and gas pressure in Fig. 5.  However, at the threshold pressure~\cite{7} for the
corona splash on a smooth surface, there is a qualitatively
different distribution of droplet sizes.  Fig. 6a shows there is no
corona and that discrete droplets emerge from the expanding liquid
which has periodic undulations along its rim. The ink spots shown in
the inset at the lower left corner of Fig. 6b have the striking
feature that they fall in a horizontal line, indicating that the ejected droplets
have the same angle between their trajectories and the substrate.  In
addition, the spots are approximately equally spaced.  The main
panel of Fig. 6b shows that the Fourier transform of the lateral
positions of the spots has a peak indicating this spatial order.
Moreover the spot sizes are more uniform than those seen at high
pressures in the inset to Fig. 5.  Fig. 6c shows $N(r)$ at pressures
close to $P_T$.  The peak at $r_0 = 0.11$ mm indicates that most
droplets are about the size of the rim undulations.
\\

\noindent\textbf{IV. Conclusions}\\

By controlling the gas pressure and surface roughness we have
identified two mechanisms for splashing: the surrounding gas is
responsible for corona splashing and surface roughness is
responsible for prompt splashing. This explains the long-standing
puzzle about why two distinct types of splashes exist.  We have also
found that there are characteristic lengths in the distribution of
ejected drops in both cases.  Similar exponential dependence is
found in ligament breakup, implying a possible connection with that
process\cite{10,11}.  However, the exponential dependence of droplet
sizes found in the splashing of liquids is in contrast to what is
found in the shattering of a solid.  In that case, there is a
power-law distribution of sizes of the shattered fragments and no
characteristic length scale\cite{12,13}.  The characteristic lengths
we find in liquid splashing experiments reveal the microscopic
length scales associated with the droplet creation process.  On
rough surfaces, the length scales we determine are consistent with
our interpretation of the prompt splash in terms of the surface
roughness.  These results also suggest a means for controlling the
sizes of ejected droplets in a splash.

\textbf{Acknowledgement} We wish to thank Qiti Guo, Priyanka Jindal,
David Qu\'{e}r\'{e}, Mathilde Callies-Reyssat,  and Wendy Zhang for
helpful discussions. This work was supported by MRSEC DMR-0213745
and NSF DMR-0652269.

\end{document}